# An Empirical Study on End-users Productivity Using Model-based Spreadsheets[*]

Laura Beckwith
HCIResearcher, Denmark
beckwith@hciResearcher.com

Jácome Cunha
Universidade do Minho, Portugal
jacome@di.uminho.pt

João Paulo Fernandes
Universidade do Minho &
Universidade do Porto, Portugal
jpaulo@{di.uminho.pt,fe.up.pt}

João Saraiva
Universidade do Minho, Portugal
jas@di.uminho.pt

**ABSTRACT**

Spreadsheets are widely used, and studies have shown that most end-user spreadsheets contain non-trivial errors. To improve end-users productivity, recent research proposes the use of a model-driven engineering approach to spreadsheets.
In this paper we conduct the first systematic empirical study to assess the effectiveness and efficiency of this approach. A set of spreadsheet end users worked with two different model-based spreadsheets, and we present and analyze here the results achieved.

## 1 INTRODUCTION

Spreadsheets can be viewed as programming environments for non-professional programmers, the so-called "end users" [Nardi, 1993]. An end user is a teacher, an engineer, a secretary, an accountant, in fact almost anyone except a trained programmer. These people use computers to get their job done; often they are not interested in programming per se. End-user programmers vastly outnumber professional ones creating every year hundreds of millions of spreadsheets [Scaffidi et al., 2005]. As numerous studies have shown, this high rate of production is accompanied by an alarming high rate of errors, with some reporting that up to 90% of real-world spreadsheets contain errors [Panko, 2000, Rajalingham et al., 2001, Powell and Baker, 2003].

In order to overcome these limitations of spreadsheets, a considerable amount of research has been recently done by the human computer interaction community [Abraham and Erwig, 2006, Cunha et al., 2010, 2009b,a, Engels and Erwig, 2005, Erwig et al., 2005]. One of the promising solutions advocates the use of a Model-Driven Engineering (MDE) approach to spreadsheets. In such an approach, a business model of the spreadsheet data is defined, and then end users are guided to introduce data that conforms to the defined model [Cunha et al., 2009b]. Indeed, several models to represent the business logic of the spreadsheet have been proposed, namely, templates [Abraham and Erwig, 2006, Erwig et al., 2005], ClassSheets [Cunha et al., 2010, Engels and Erwig, 2005], relational models [Cunha et al., 2009a]. Several techniques to infer such models from a (legacy) spreadsheet data have also been studied [Abraham and Erwig, 2006, Cunha et al., 2010].

[*] Supported by Fundação para a Ciência e a Tecnologia, grants no. SFRH/BPD/73358/2010, SFRH/BPD/46987/2008 and PTDC/EIA-CCO/108613/2008.

Although all these works claim that a MDE approach improves end-users productivity, there is no detailed evaluation that supports this idea besides our first attempt in [Cunha et al., 2011]. In this paper, we present a complete empirical study that we have conducted with the aim of analyzing the influence of using models in end-users spreadsheet productivity. In this study we consider two different model-based spreadsheets, as proposed in [Cunha et al., 2009a,b]. We assess end-users productivity in introducing, updating and querying data in those two model-based spreadsheets and in a traditional one. As the models we consider represent database-like spreadsheets only it should be clear that we are not analyzing all possible (types of) spreadsheets. Nevertheless, even considering spreadsheets strongly related to databases, the domain of our tests is clearly the spreadsheet environment.

In this paper we wish to answer the following research questions:

**RQ1** Do end users introduce fewer errors when they use one of the model-based spreadsheet versus the original unmodified spreadsheet?

**RQ2** Are end users more efficient using the model-based ones?

**RQ3** Do particular models lead to fewer errors in particular tasks?

The study we conducted to answer these questions is necessary and useful, since it is based on a sound experimental setting and thus allow us to draw sound conclusions for further studies on how to improve spreadsheet end users productivity.

This paper is structured as follows: in Section 2 we present the model-based spreadsheets we considered in our study. In Section 3 we describe the design of our study. We present and analyze in detail the results of our study in Section 4. Several threats to validity are discussed in Section 5 and in Section 6 we draw our conclusions.

## 2 MODEL-BASED SPREADSHEETS

There have been proposed two different techniques to tackle the problem of preventing errors in spreadsheets [Cunha et al., 2009b,a]. In order to introduce these works we will rely on the spreadsheet shown in Figure 1. This spreadsheet represents a movie renting system registering movies, renters and rents. Labels in the spreadsheet should be self-explicative.

| | A | B | C | D | E | F | G | H | I | J | K | L |
|---|---|---|---|---|---|---|---|---|---|---|---|---|
| 1 | movieID | title | year | director | language | renterNr | renterNm | renterPhone | rentStart | rentFinished | rent | totalToPay |
| 2 | mv23 | Little Man | 2006 | Keenen Wayans | English | c33 | Paul | 3334433 | 01-04-2010 | 26-04-2010 | 0,5 | 12,50 |
| 3 | mv1 | The OH in Ohio | 2005 | Billy Kent | English | c33 | Paul | 3334433 | 30-03-2010 | 23-04-2010 | 0,5 | 12,00 |
| 4 | mv21 | Edmond | 2005 | Stuart Gordon | English | c26 | Smith | 4445467 | 02-04-2010 | 04-04-2010 | 0,5 | 1,00 |
| 5 | mv102 | You, Me and D. | 2001 | Anthony Russo | English | c3 | Michael | 5551212 | 22-03-2010 | 03-04-2010 | 0,3 | 3,60 |
| 5 | mv102 | You, Me and D. | 2001 | Anthony Russo | English | c3 | Michael | 5551212 | 22-03-2010 | 03-04-2010 | 0,3 | 3,60 |

Figure 1: Part of a spreadsheet representing a movie renting system.

### 2.1 The Refactored Spreadsheet Model

The spreadsheet shown in Figure 1 defines a valid model to represent the information of the renting system. However, it contains redundant information. For example, the information about the client **Paul** appears four times in the spreadsheet! This kind of redundancy makes the maintenance of the spreadsheet complex and error-prone, specially for end users. A mistake is easily made, for example, by mistyping a name and thus corrupting the data.

The same information can be stored without redundancy. In fact, in the database community, techniques for database normalization are commonly used to minimize duplication of information and improve data integrity. Database normalization is based on the detection and exploitation of functional dependencies inherent in the data [Codd, 1970]. We have adapted these techniques to work with spreadsheets: from the spreadsheet data we infer a set of normalized FDs, and from them, we compute a relational model [Cunha et al., 2009a]. A spreadsheet respecting such model is shown in Fig. 2.

Figure 2: Part of a refactored spreadsheet representing a movie renting system.

The obtained modularity solves two well-known problems in databases, namely update and deletion anomalies [Codd, 1970]. The former problem occurs when we change information in one tuple but leave the same information unchanged in the others. In our example, this may happen if the user changes the rent per day of movie number mv23 from 0.5 to 0.6. In the modular spreadsheet that value occurs only once in the movie table and so that problem will never occur. The latter problem happens when we delete some tuple and lose other information as a side effect. For example, if the user deletes row 3 in the original spreadsheet all the information about movie mv1 is eliminated. Since we have a deep knowledge about the relations and relationships in the data, we can generate spreadsheets that respect them. For example, in the renter table, the generated spreadsheet will not allow the user to introduce two renters with the same number (renterNr). If that error occurs the spreadsheet system should warn the user as shown in Figure 3. Obviously, it is not possible to perform this validation in the original spreadsheet. The refactored spreadsheet not only improves modularity and detects the introduction of incorrect data, it also eliminates redundancy: the redundancy present in the original spreadsheet has been eliminated. As expected, the information about renters (and movies) occurs only once. These features should help end users to commit less errors. In the sequel, this model will be referred as refactored.

Figure 3: Introducing a new row with a previously used code produces an error.

### 2.2 The Visual Spreadsheet Model

In [Cunha et al., 2009b], the authors proposed a technique to enhance a spreadsheet system with mechanisms to guide end users to introduce correct data. Using the relational database schema induced by the data we construct a spreadsheet environment that respects that schema. For example, for the movie spreadsheet, the system does not allow end users to introduce two different movies with the same number (movieID). Instead, it offers to the user a list of possible movies, such that he can choose the value to fill in the cell. This new spreadsheet, that we show in Figure 4, also includes advanced features which provide information to the end user about correct data that can be introduced.

Figure 4: Part of a visual spreadsheet representing a movie renting system.

We consider 3 types of advanced features. First, we consider bidirectional auto-completion of column values: based on the relational schema, we know that some columns (consequents) depend on another column (antecedents). Both antecedent and consequent columns have combo boxes that allow users to choose values instead of writing them. Using this knowledge we created a mechanism that automatically fills in consequent columns when antecedent columns are filled in. When values are written in consequent columns, the values in the antecedent columns are filtered, showing only the ones that can imply the chosen consequents. Using the bidirectional auto-completion feature the spreadsheet system guarantees that the user does not introduce data that violates the inferred model. The second feature is non-editable columns: this feature prevents the user from editing consequent columns since this could break the relationship with the antecedents. Note that, such columns are automatically filled in by selecting the corresponding antecedent. Finally, we consider safe deletion of rows: the user receives a warning when deleting of a row provokes the deletion of data not represented elsewhere.

Like in modern programming environments, the refactored spreadsheet system also offers the possibility of using traditional editing, i.e. the introduction of data by editing each of the columns. When using traditional editing the end user is able to introduce data that violates the model inferred from the previous spreadsheet data. The spreadsheet environment includes a mechanism to re-calculate the relational database model after traditional editing. This new relational model is used to guide the end user in future editing. From now on, this model of spreadsheet will be referred as the visual model.

## 3 STUDY DESIGN

As suggested in [Perry et al., 2000] we organized the study as follows:

1. Formulating hypothesis to test: we spent a considerable amount of time organizing our ideas and finally formulating the hypothesis presented in this work: model-based spreadsheets can help end users committing less errors editing and querying spreadsheets.

2. Observing a situation: once we got enough and appropriate qualified participants we ran the study itself. During the study, we screen casted the participants' computers and afterwards we collected the spreadsheets they worked on.

3. Abstracting observations into data: we computed a series of statistics, that we present in detail in Section 4, over the spreadsheets participants developed during the study: we graded their performance and measured the time they took to perform the proposed tasks. All the data we used is available at the SSaaPP project web page *http://ssaapp.di.uminho.pt*. The tasks and spreadsheets participants received are also available.

4. Analyzing the data: the enormous collection of data that we gathered was later systematically analyzed. This analysis is also presented in this paper, in Section 4.

5. Drawing conclusions with respect to the tested hypothesis: based on the results we obtained, we finally drawn some conclusions. We were also able to suggest some future research paths based on our work, which are presented in the Section 6.

Our study aimed to answer if participants were able to perform their tasks with more accuracy and/or faster given the experimental environments. We used a within subjects design, where each participant received 3 spreadsheets, one for each problem (DISHES, PROPERTIES, PROJECTS). Each of the 3 spreadsheets was randomly distributed under one of the 3 model (original, visual, refactored). Participants were asked to do various tasks in each spreadsheet: data entry, editing, and calculations. They were encouraged to work as quickly, but were not given time limits.

### 3.1 Methodology

Participants started the study by filling out a background questionnaire so we could collect their area of study and previous experience with spreadsheets, other programming languages and English comfort (Portuguese is their mother language). An introduction to the study was given orally in English, this was explicitly not a tutorial for the different environments because the goal was to see if even without any introduction to the various models the participants would still be able to understand and complete the tasks. The participants were asked to work as quickly and accurately as possible. Since the order of the spreadsheets was randomized, they were told that the other sitting around them might appear to be moving faster, but that some tasks were shorter than others. After 2 hours participants were stopped if they were not already finished. Following the tasks they had a post session questionnaire which contained questions assessing their understanding of the different models, (3 questions for refactored and 4 for visual).

Correct answers could only be given by participants having understood the running models. Grading the questionnaires was done as follows: a correct answer receives total points; an incorrect answer receives 0 points and an answer that is not incorrect nor (totally) correct receives half of the points. We recorded the users screens using screen capture technology. At the end of the study the users completed spreadsheets were saved and graded for later analysis.

### 3.2 Participants

Recruitment was conducted through a general email message to the university, asking for students with spreadsheet experience and comfort with English. Of the hundreds that responded (here was a compensation involved), participants were selected based on spreadsheet experience, comfort with English, and majors outside of computer science and engineering. In total, 38 participants finished the study with data we were able to use (25 females, 11 males, and 2 who did not answer about their gender). Two participants did not try to solve one of the proposed tasks; for these participants, we included in the study only the tasks they undertook. A few participants' machines crashed and therefore they were eliminated from the study. The majority of participants were between 20-29 years of age, with the remaining under 20. All were students at the university. About 2/3 were working on their Baccalaureate degree, the remaining on their Masters. None were studying computer science or engineering and the most represented majors were medicine, economics, nursing and biology. A variety that is good for representing the end-user population of spreadsheets.

## 3.3 Tasks

The task lists were designed to include tasks that are known to be problematic in spreadsheets, which involve data insertion, edition and the use of formulas. The tasks were 1) add new information to the spreadsheet, 2) edit existing data in the spreadsheets and 3) do some calculations using the data in the spreadsheets.

Some of the tasks asked users to add many new rows of data, with the aim of a repetitive task being common in real-world situations. As we were designing the tasks, we imagined a type of data entry office scenario, where an office worker might receive on paper data which was initially filled out on a paper form and needed to be entered into a spreadsheet. This first task of data entry, in theory, should be fastest (and done with fewest entry errors) in the refactored spreadsheet. The second task, of making changes to existing data in a spreadsheet should also be easier within a refactored spreadsheet, since the change only needs to be made in one location, and therefore there would be less chance of forgetting to change it. The final task was to do some calculations using the data in the spreadsheet, such as averages, etc. This task was added because of the frequency of problems with formulas. One of the spreadsheets used in the study, PROJECTS, stores information about a house renting system (adapted from [Connolly and Begg, 2001]). This spreadsheet has information about renters, houses and their owners as well as the dates and prices of the rents.

A second spreadsheet, DISHES, contains information about sells of detergents to dish washers. Information about the detergents, prices and the stores where they are sold is present on this spreadsheets (adapted from [Powell and Baker, 2003]).

The last spreadsheet, PROJECTS, stores information about projects, like the manager and delivery date, employees instruments used (adapted from [Alhajj, 2003]).

In the task list for DISHES, 67% (39 out of 58 cells needed to be changed) of the tasks consist of inserting new data, 21% (12/58) are editing tasks and 12% (7/58) involve calculations over the data in the spreadsheet. In the task list for PROJECTS, 80% (221/277) of the tasks are for inserting new data, 7% (20/277) for edition and 13% (36/227) for calculations. Finally, for PROJECTS, inserting data tasks are 56% (64/115) of the total, whereas data editing and calculation tasks are 19% (22/115) and 25% (29/115) of the total, respectively. Grading the participants' performance was done as follows. For tasks involving adding new data to the spreadsheet or performing calculations over spreadsheet data, whenever a participant executes a task as we asked him/her to, he/she is awarded 100% of the total score for that task; on the contrary, if the participant does not at all try to solve a particular task, he/she gets no credit for that. An intermediate situation occurs when participants try to solve a task, but fail to successfully conclude it in its entirety. In this case, the participant is awarded 50% of the score for that task. For tasks involving editing data, a value in the interval 0%−100% is awarded according to the participants' success rate in such tasks. Table 1 shows the number of participants that worked on each spreadsheet and each model. Note that the distribution of models and spreadsheets by the participants is homogeneous.

|            | *original* | *refactored* | *visual* | Total |
|------------|------------|--------------|----------|-------|
| DISHES     | 12         | 13           | 12       | 37    |
| PROJECTS   | 11         | 13           | 13       | 37    |
| PROPERTIES | 14         | 11           | 13       | 38    |
| Total      | 37         | 37           | 38       |       |

Table 1: Participants per spreadsheet/model.

## 4 ANALYZING END-USER PERFORMANCE

We divide the presentation of our empiric results under two main axes: effectiveness and efficiency. In studying effectiveness we want to compare the three running models for the percentage of correct tasks that participants produced in each one. In studying efficiency we wish to compare the time that participants took to execute their assigned tasks in each of the different models. We start by effectiveness.

### 4.1 Effectiveness

Each participant was handed 3 different lists of tasks (insert, edit and query data) to perform on 3 different spreadsheets (DISHES, PROJECTS, PROJECTS). Each spreadsheet, for the same participant, was constructed under a different model (original, refactored visual).

For each spreadsheet, and for each model, we started by analyzing the average of the scores obtained by participants. We shown in Figure 5 the results of such analysis.

|            | original | refactored | visual |
|------------|----------|------------|--------|
| DISHES     | 86%      | 76%        | 78%    |
| PROJECTS   | 73%      | 68%        | 78%    |
| PROPERTIES | 75%      | 64%        | 62%    |

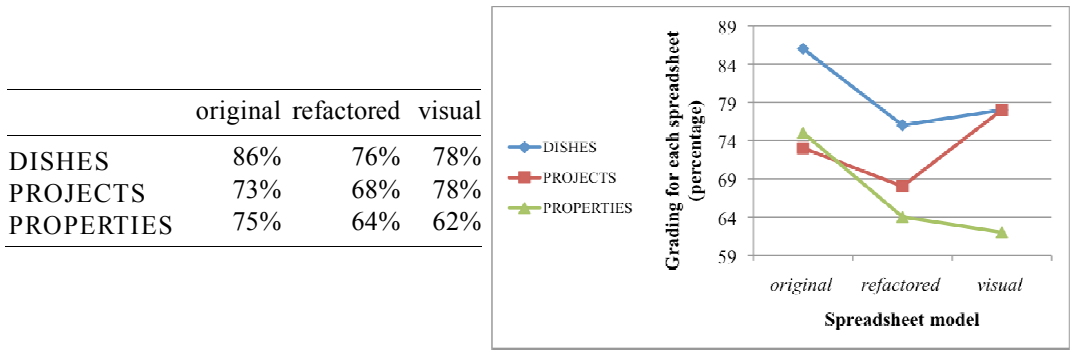

Figure 5: Global effectiveness results.

We notice that no spreadsheet model is the best for all spreadsheets in terms of effectiveness. Indeed, we may even notice that spreadsheets in the traditional style, the original model, turned out to be the best for both the DISHES and PROJECTS spreadsheets. The visual model suited the best for the PROJECTS spreadsheet.
In the same line of reasoning, there is no worst model: refactored achieved the worst results for DISHES and PROJECTS; visual got the lowest average scores for PROJECTS.

Nevertheless, these results seem to indicate that the models that we have developed are not effective in reducing the number of errors in spreadsheets, since one of them is always the model getting the lowest scores. This first intuition, however, deserves further development. For once, on the theoretical side, one may argue that original is, without a doubt, the model that end users are accustomed to. Recall that in the study, we opted to leave out participants with computer science backgrounds, who could be more sensible to the more complex models refactored and visual, preferring to investigate such models on traditional users of spreadsheets. On the other hand, we remark that these more complex models were not introduced; a part of our study was also to learn whether or not they could live on their own.

Our next step was to investigate whether the (apparent) poor results obtained by complex models are due to their own nature or if they result from participants not having understood them. So, we studied participations that did not achieve at least 50%, which are distributed by the spreadsheet models as follows: original, 0%, refactored, 25% and visual, 21%. While in original no participation was graded under 50%, this was not the case for refactored and visual, which may have degraded their overall average results.

For these participations, we analyzed the questionnaire that participants were asked to fill in after the session. The average classifications for the post session questionnaires, for participations that were graded under 50% is 24% for refactored and 31% for visual.
These results show that participants obtaining poor gradings on their effectiveness, also got poor gradings for their answers to the questions assessing how they understood the models they had worked with. In fact, such participants were not able to answer correctly to (at least) two thirds of the questions raised in the post session questionnaire. From such results we can read that 1/4 of participants was not able to understand the more complex models, which might have caused a degradation of the global effectiveness results for these models. This also suggests that if these models are to be used within an organization, it is necessary to take some time to introduce them to end users in order to achieve maximum effectiveness. Nevertheless, even without this introduction, the results show that the models are competitive in terms of effectiveness: at most they are 13% worst than the original model, and for one of the spreadsheets, the visual model even got the best global effectiveness.

### 4.1.1 Effectiveness by Task Type

Next, we wanted to realize how effective models are to perform each of the different types of tasks that we have proposed to participants: data insertion, edition and statistics.

*i) Data insertion*: The results presented in Figure 6 show, for each model, how effective participants were in adding new information to the spreadsheets they received.

|  | original | refactored | visual |
|---|---|---|---|
| DISHES | 91% | 90% | 81% |
| PROJECTS | 76% | 60% | 75% |
| PROPERTIES | 86% | 67% | 68% |

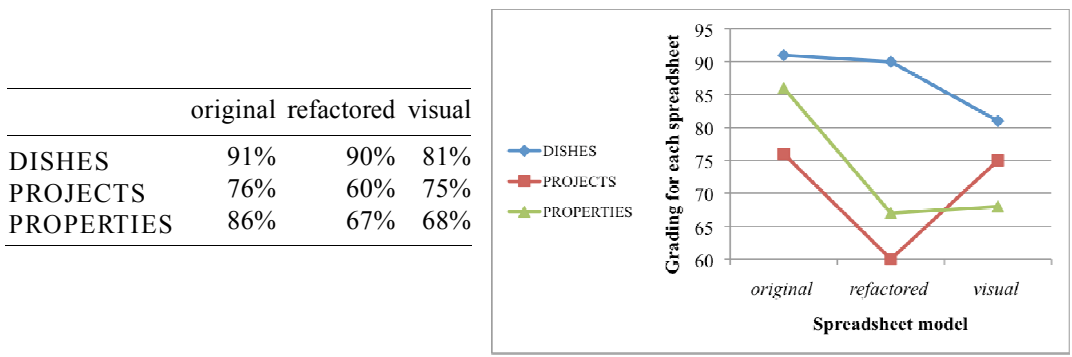

Figure 6: Effectiveness results for data insertion.

The original model revealed to be the most effective, for all three spreadsheets, being closely followed by refactored and visual for DISHES, and by visual for PROJECTS. The refactored model, for PROJECTS, and the models refactored and visual, for PROJECTS, proved not to be competitive for data insertion, in the context of the study.

Again, we believe that this in part due to these models not having been introduced previously to the study: the insertion of new data is the task that is most likely to benefit from totally understanding of the running model, and also the one that can be otherwise most affected. This is confirmed by the effectiveness results observed for other task types, that we present next.

*ii) Data edition*: Now, we analyze the effectiveness of the models for editing spreadsheet data. The results presented in Figure 7 show that once a spreadsheet is populated, we can effectively use the models to edit its data.

|  | original | refactored | visual |
|---|---|---|---|
| DISHES | 91% | 82% | 82% |
| PROJECTS | 54% | 62% | 50% |
| PROPERTIES | 65% | 98% | 48% |

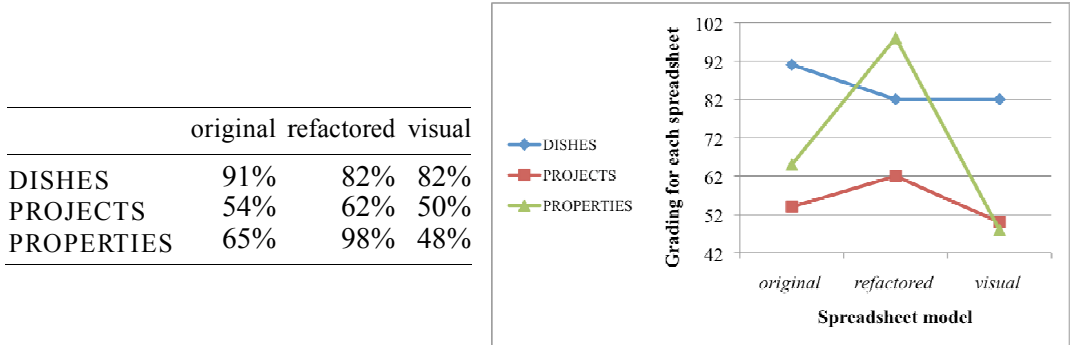

Figure 7: Effectiveness results for data edition.

This is the case of refactored for PROJECTS and specially for PROJECTS. original is the most effective in data editing for DISHES. visual is comparable to refactored for DISHES, but for all other spreadsheets, it always achieves the lowest scores among the three models.

*iii) Statistics*: Finally, we have measured the effectiveness of the models for performing calculations over spreadsheet data, obtaining the results shown in Figure 8.

|  | original | refactored | visual |
|---|---|---|---|
| DISHES | 52% | 37% | 57% |
| PROJECTS | 19% | 76% | 13% |
| PROPERTIES | 44% | 57% | 51% |

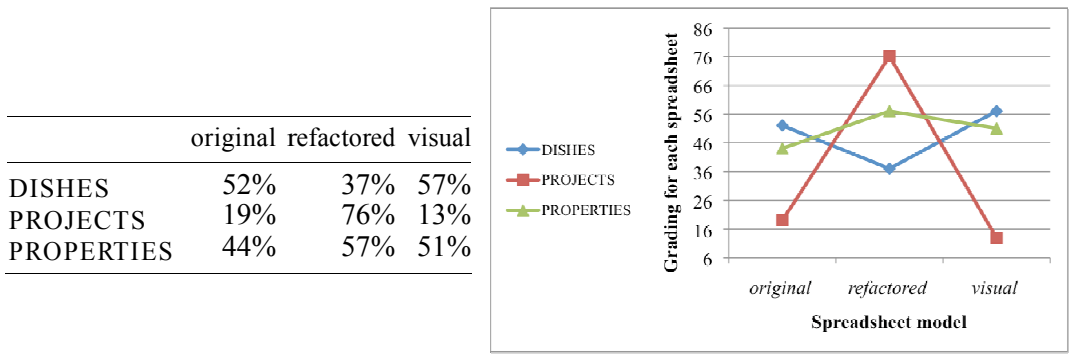

Figure 8: Effectiveness results for statistical calculations.

We can see that visual obtained the best results for DISHES, and that refactored obtained the best results for both spreadsheets PROJECTS and PROJECTS. We can also see that all models obtained the worst results for exactly one spreadsheet.

Results from i), ii) and iii) confirm that the models are competitive. On the other hand, these results allow us to draw some new conclusions: if the models are going to be used within an organization, it may not always be necessary to introduce them prior to their use. Indeed, if an organization mostly edits spreadsheet data or computes new values from such data, and does not insert new data, then the models, and specially refactored, may deliver good results even without being explained. These results also show that it is inserting data that models need to be better understood by end users in order to increase effectiveness.

## 4.2 Efficiency

In this section, we analyze the efficiency results obtained in our study by the models that we have been considering in this paper.

We started by measuring, for each participant, and for each spreadsheet, the time elapsed from the moment participants started reading the list of tasks to undertake until the moment they completed the tasks proposed for that particular spreadsheet and moved on to a different spreadsheet or concluded the study. We are able of calculating these times by looking at the individual screen activity that was recorded during the study, for each participant: the participant stopping interacting with the computer signals the end of his/her work on a spreadsheet. The measured period therefore includes the time that participants took trying to understand the models they received each spreadsheet in.

Figure 9 presents the average of the overall times, for each spreadsheet and for each model. We can see that refactored and visual are competitive in terms of efficiency: participants performed fastest for DISHES in visual, and fastest, by a marginal factor, for the PROPERTIES in refactored. original got the best efficiency measurements for PROJECTS, also by a marginal factor. Again, note that no introduction to these models preceded the study.

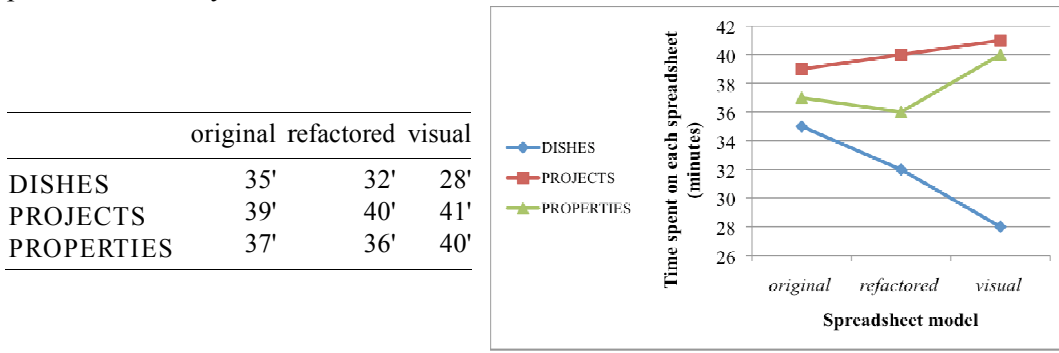

|            | original | refactored | visual |
|------------|----------|------------|--------|
| DISHES     | 35'      | 32'        | 28'    |
| PROJECTS   | 39'      | 40'        | 41'    |
| PROPERTIES | 37'      | 36'        | 40'    |

Figure 9: Global efficiency results.

Therefore, it is reasonable to assume that the results in Figure 9 include some time overhead. In an attempt to measure this overhead, which is a consequence of participants having to analyze a new model, we extracted some information out of the participants' screen activity. Indeed, we measured the time elapsed from the moment participants started reading, for each spreadsheet, the list of tasks to perform, until the moment they actually began editing the spreadsheet. We assume that this period corresponds exactly to the overhead of understanding each model (obviously increased by the time spent reading the list of tasks, which we are not able of isolating further, but that should be constant for any spreadsheet model, since the task list does not change with the model). These results are presented in Table 2.

|  | original | refactored | visual |
|---|---|---|---|
| DISHES | 35' | 32' | 28' |
| PROJECTS | 39' | 40' | 41' |
| PROPERTIES | 37' | 36' | 40' |

Table 2: Average overhead results.

We notice that there is a constant average overhead of 2 minutes for almost all models and spreadsheets, with the most significant exceptions occurring for refactored, for both DISHES and PROJECTS. In these cases, we can clearly notice an important time gap, which provides some evidence that refactored is most likely the hardest model to understand. This also comes in line with previous indications that the merits of models can be maximized if we explain them to end users. For the particular case of efficiency, this means that the results shown in Figure 9 could be further improved for the more complex models, and particularly for refactored.

## 5 THREATS TO VALIDITY

As suggested by Perry et al. [Perry et al., 2000], we discuss three types of influences that might limit the validity of our study.

Construct Validity: Do the variables and hypotheses of our study accurately model the research questions?

i) Measuring the time overhead: when studying efficiency, we measured the overhead of understanding each model as the period of time that participants stopped interacting with a spreadsheet and started editing the next one. In this period, it might have been the case that participants, instead of being focused on understanding the new model, took the time to do something else, like resting. This could affect our conclusions in terms of efficiency. However, during the study, participants where supervised by two authors, who observed that this was not the case. Even if we were not able to spot a small number of such occurrences, the differences in the results should be minimal and so they should not affect our conclusions.

ii) Original model: In our study, we have used three spreadsheets that we have assumed to be in the original model. What we are saying is that these three spreadsheets are representative of the spreadsheets normally defined by end users. Although this set of spreadsheets may be too large to be represented by (any) three spreadsheets, we have taken DISHES, PROJECTS and PROJECTS directly, or with small changes, from other works on general purpose spreadsheets [Alhajj, 2003, Connolly and Begg, 2001, Powell and Baker, 2003].

Internal Validity: Can changes in the dependent variables be safely attributed to changes in the independent variable?

i) Accuracy of the analysis: Some of the inferences we make in this paper deserve further analysis. To some extent, we assume that our models could achieve better results if a tutorial has been given to the participants. In fact, we have no proof of this, but the evidences from the study seem to strongly indicate this fact. A new study is required to prove this, though.

ii) Accuracy of measurements: Each task proposed to participants was individually graded. For most of the cases, this was done automatically using OpenOffice scripts. These scripts and their results were tested and checked. The cases for which an automatic grading was not possible were carefully graded by hand. All grades were validated by two authors and were randomly re-checked. Since we have more than 1400 grades, it is virtually impossible to guarantee full accuracy. This could affect the results observed for dependent variables (efficiency and effectiveness) without really the independent variables (the models considered) having changed. Nevertheless, if imprecisions exist in the grades, they should be equally distributed by the 3 models and thus they should not affect the overall results.

The measurement of times that lead to the results presented earlier was achieved by visualizing the screen casts made during the study for. Being a manual and repetitive task, it is subject to imprecisions. Also, not being able to visualize the actual participants' behavior now may lead to imprecise measurements. We are confident that, even if there are imprecisions, such imprecisions should be distributed evenly by all measurements and thus do not influence the efficiency results or the conclusions that we draw based on them.

External Validity: Can the study results be generalized to settings outside the study?

i) Generalization: In this study we used three different spreadsheets from different domains. We believe that the results can be generalized to other spreadsheets, although probably not to all. The models we developed are not restricted to any particular spreadsheet, and thus, the results should be the same if the study was run with a different set of spreadsheets.

ii) Industrial usage: Participants were asked to simulate industrial activity: they received some data on paper that they had to register in a spreadsheet. Although we have tried to create a realistic environment for the study, it is likely that people would respond differently in an industrial context. Also, participants were University students whose technical abilities and experience surely differ from other spreadsheet users. Nevertheless, we believe that this affects no spreadsheet/model in particular. Possible impacts would affect all spreadsheets/models in the same way and thus the overall results apply. We believe that if the study was conducted on an industrial environment, the conclusions should be similar.

## 6 CONCLUSIONS

In this paper, we have presented the results of an empirical study that we conducted in order to assess the practical interest of models for spreadsheets.

According to [Perry et al., 2000], three topics deserve further analysis. The first is accuracy of interpretation: this study was prepared carefully and a significantly large number of end users participated in it. Our goal here was to guarantee that the results are not unknowingly influenced. For this, it also contributes the fact that we make all the elements of this study available, both in this paper and online. The second topic is relevance: MDE is one of the most significant research areas in software engineering. We adapted some techniques from this field to spreadsheets and showed that they can bring benefit not only for professional users but also for end users. The last topic is impact: our first results show that MDE can bring benefits for spreadsheet end users. This is a promising research direction, that we believe can be further explored, particularly in contexts similar to the one of this paper. From the preparation of the study, from running it and from its results, we can summarize our main contributions as follows: we have shown that MDE techniques can be adapted for end-users software; moreover, we provided empirical evidence that models can bring benefits to spreadsheet end users; finally, we have proposed a methodology that can be reused in studies similar to the one we have conducted.

Finally, we seek to answer the research questions that we presented in the introduction of this paper, which correspond exactly to the questions our study was designed to answer.

**RQ1**: Our observations indicate that there is potential for improving end-user effectiveness using model-based spreadsheets. Even if this is not always the case, our results also indicate that deeper insight on the spreadsheet models is required to maximize effectiveness. Indeed, we believe that the effectiveness results for refactored and visual could have been significantly better if these models had been preliminary presented to the participants of our study.

**RQ2**: We observed that, frequently, the more elaborate spreadsheet models allowed users to perform faster. Nevertheless, we were not fully able of isolating the time that participants took trying to understand the models they were given. So, we believe that the observed efficiency results could also be better for refactored and visual if they had been introduced.

**RQ3**: Although this was not observed for inserting tasks, the fact is that, for editing and querying data the models did help end users. Furthermore, the results seem to indicate that the inserting data task is the one that benefits the most from better understanding the models.
With this study we have shown that there is potential in MDE techniques for helping spreadsheet end users. The study of these techniques for professional users of spreadsheets seems a promising research topic. Moreover, the use of MDE techniques in other non-professional softwares should also be investigated.

## REFERENCES

Robin Abraham and Martin Erwig. Inferring templates from spreadsheets. In Proc. of the 28th Int. Conference on Software Engineering, pages 182–191, New York, NY, USA, 2006. ACM. ISBN 1-59593-375-1.

Reda Alhajj. Extracting the extended entity-relationship model from a legacy relational database. Information Systems, 28(6):597–618, 2003. ISSN 0306-4379. doi: http://dx.doi.org/10.1016/S0306-4379(02)00042-X.

E. F. Codd. A relational model of data for large shared data banks. Communications of the ACM, 13:377–387, June 1970. ISSN 0001-0782. doi: http://doi.acm.org/10.1145/362384.362685.

Thomas M. Connolly and Carolyn Begg. Database Systems: A Practical Approach to Design, Implementation, and Management. Addison-Wesley Longman Publishing Co., Inc., Boston, USA, 2001. ISBN 0201708574.

J. Cunha, J. Saraiva, and J. Visser. From spreadsheets to relational databases and back. In PEPM'09: Proceedings of the 2009 ACM SIGPLAN Workshop on Partial Evaluation and Program Manipulation, pages 179–188, New York, NY, USA, 2009a. ACM. ISBN 978-1-60558-327-3.

J. Cunha, J. Saraiva, and J. Visser. Discovery-based edit assistance for spreadsheets. In Proc. of the 2009 IEEE Symposium on Visual Languages and Human-Centric Computing, pages 233–237, Washington, DC, USA, 2009b. IEEE Computer Society. ISBN 978-1-4244-4876-0.

J. Cunha, M. Erwig, and J. Saraiva. Automatically inferring ClassSheet models from spreadsheets. In Proc. of the 2010 IEEE Symposium on Visual Languages and Human-Centric Computing, pages 93–100, Washington, DC, USA, 2010. IEEE Computer Society.

J. Cunha, L. Beckwith, J. Paulo Fernandes, and J. Saraiva. End-users productivity in model-based spreadsheets: An empirical study. In IS-EUD '11: Third International Symposium on End-User Development, 2011. to appear.

Gregor Engels and Martin Erwig. ClassSheets: Automatic generation of spreadsheet applications from object-oriented specifications. In Proceedings of the 20th IEEE/ACM International Conference on Automated Software Engineering, pages 124–133, New York, NY, USA, 2005. ACM. ISBN 1-59593-993-4.

Martin Erwig, Robin Abraham, Irene Cooperstein, and Steve Kollmansberger. Automatic generation and maintenance of correct spreadsheets. In ICSE '05: Proceedings of the 27th International Conference on Software Engineering, pages 136–145, New York, NY, USA, 2005. ACM. ISBN 1-59593-963-2.

Bonnie A. Nardi. A Small Matter of Programming: Perspectives on End User Computing. MIT Press, Cambridge, MA, USA, 1993. ISBN 0262140535.

Raymond R. Panko. Spreadsheet errors: What we know. What we think we can do. Proceedings of the Spreadsheet Risk Symposium, European Spreadsheet Risks Interest Group, July 2000.

Dewayne E. Perry, Adam A. Porter, and Lawrence G. Votta. Empirical studies of software engineering: A roadmap. In ICSE '00: Proceedings of the Conference on The Future of Software Engineering, pages 345–355, New York, NY, USA, 2000. ACM. ISBN 1-58113-253-0.

Stephen G. Powell and Kenneth R. Baker. The Art of Modeling with Spreadsheets. John Wiley & Sons, Inc., New York, NY, USA, 2003. ISBN 0471209376.

Kamalasen Rajalingham, David Chadwick, and Brian Knight. Classification of spreadsheet errors. European Spreadsheet Risks Interest Group (EuSpRIG), 2001.

Christopher Scaffidi, Mary Shaw, and Brad Myers. Estimating the numbers of end users and end user programmers. In Proceedings of the 2005 IEEE Symposium on Visual Languages and Human-Centric Computing, pages 207–214, Washington, DC, USA, 2005. IEEE Computer Society. ISBN 0-7695-2443-5.